# A Universal Basic Income For Brazil: Fiscal and Distributional Effects of Alternative Schemes

**Rozane Bezerra de Siqueira[1*], José Ricardo Bezerra Nogueira[1*]**

1 Department of Economics, Centro de Ciências Sociais Aplicadas, Universidade Federal de Pernambuco, Av. dos Economistas s/n, Recife-PE, CEP 50740-580

* For correspondence: rozane_siqueira@yahoo.com.br; jrbnogueira@yahoo.com.br

(July 2021)[1]

**ABSTRACT**: The outbreak of the Covid-19 pandemic has led to an increasing interest in Universal Basic Income (UBI) proposals as it exposed the inadequacy of traditional welfare systems to provide basic financial security to a large share of the population. In this paper, we use a static tax-benefit microsimulation model to analyse the fiscal and distributional effects of the hypothetical implementation in Brazil of alternative UBI schemes which partially replace the existing tax-transfer system. The results indicate that the introduction of a UBI/Flat Tax system in the country could be both extremely effective in reducing poverty and inequality and economically viable.

## 1. Introduction

Over the past few years, interest in Universal Basic Income (UBI) proposals has grown enormously across the world. Gentilini, Grosh and Yemtsov (2020), in a chapter of a book on UBI published by the World Bank (Gentilini, Grosh, Rigolini and Yemtsov, 2020), remark that over the last decade alone 91 books were published on the subject, and several pilot programs were implemented around the world. More recently, the outbreak of the Covid-19 pandemic substantially increased the interest on UBI proposals, as it exposed the inadequacy of traditional welfare systems to provide financial security to large segments of the labour market, particularly self-employed workers and small business.

A UBI is usually conceived of as a regular fixed amount of money provided by the state to every citizen, regardless of income level, employment status or any other conditionality. The only limiting condition permitted would be the recipient's age. The notion dates back to the 18th century and was introduced by Thomas Paine (1797). In the second half of the last century interest in the idea was renewed with James Buchanan (1960, 1967), Milton Friedman (1962), and James Tobin (1967). In the 1990's a new wave of academic work on UBI was initiated with key authors such as Philippe van Parijs (1992), James Meade (1993), and Anthony Atkinson (1995).[2]

---

[1] Version prepared for presentation at the 22nd Basic Income Earth Network Congress, 18th-21st August 2021.

The upsurge of interest in UBI proposals in the last decade emerged initially in advanced countries, motivated to a great extent by growing inequality and rising job insecurity associated with technological progress (particularly, automation) and globalization. But proposals also proliferated among developing countries, where the limitations of the existing social protection systems to reducing extreme poverty and inequality tend to figure prominently among the motivations for implementing a UBI.[3]

With the Covid-19 pandemic, awareness of the fundamental role a UBI may have in such a context is increasing among both developed and developing countries. As a response to the dramatic negative impact of the pandemic on the economic circumstances of households, several countries have enacted emergency cash transfer programs. In Brazil, more than 70 million individuals, corresponding to more than 40% of the working age population, have applied to receive the emergency cash benefit created by the federal government in order to mitigate the effects of household impoverishment during the pandemic crises.

Notwithstanding the existence, since 2004, of a law which establishes a UBI to be progressively implemented in the country, the so-called *Lei de Renda Básica de Cidadania* (Citizen´s Basic Income Law), until the coronavirus pandemic no movement had been made in this direction. This law was proposed by Eduardo Suplicy, a federal senator in the Brazilian parliament from 1991 to 2015, who campaigns for the introduction of a UBI in Brazil since the early 1990s.[4] However, since the enactment of the law till recently, the political debate on the subject had faded. A major reason for this is the belief that a UBI is not fiscally sustainable in the long run and that the existing conditional cash transfer programs are more adequate to fight poverty and redistribute income.[5]

The purpose of this study is to use a static tax-benefit microsimulation model to analyse the fiscal and distributional implications of the hypothetical implementation in Brazil of three alternative UBI schemes which partially replace the existing tax-transfer system. Two of the schemes considered are versions of the so-called 'basic income/flat tax proposal', which combines a universal transfer with a single-rate tax on all other incomes. The third scheme allows for a progressive tax structure.

---

[2] For a full review of the origin, developments, and key issues concerning UBI, see Van Parijs and Vanderborght (2017) and Gentilini et al. (2020).

[3] Hoynes (2019) provides a discussion on the implementation of UBI in advanced countries, while Banerjee, Niehaus, and Suri (2019) discuss UBI in the context of developing countries.

[4] Suplicy (2013) describes the author's proposal and engagement for a UBI in Brazil.

[5] For a more detailed discussion of issues preventing the practical implementation of the *Lei de Renda Básica de Cidadania*, see Lavinas (2013).

Given the inequitable and fragmented nature of the Brazilian social protection system, one can expect, beforehand, that a move to a UBI would improve welfare. Despite a high level of spending on social protection, Brazil is one of the most unequal countries in the world and extreme poverty is still widespread, particularly among children. A study published by the Brazilian Ministry of Economy (MF, 2017) reports, for instance, that in 2015 public cash transfers represented 23% of total household income, but the absolute amount paid to the richest 20% of the population was ten times as high as that paid to the 20% poorest.[6] To a great extent this reflects the country's highly dualistic benefit structure, with generous pension regimes and relatively low provisions to poor families with children.

In this context, a UBI reform has the potential to deliver significant equity improvements at a sustainable financial cost, as the government can (in principle) offset a substantial proportion of UBI's gross cost by adjusting existing benefits' levels downwards. Additionally, UBI schemes such as those considered in this study have some advantages over the current tax-transfer system which are related to their universal, simple and transparent nature. This includes the reduction in bureaucratic costs and the minimization of opportunities for manipulation of the system by vested interests, as well as the promotion of a sense of citizenship and social cohesion.

To the best of our knowledge, there are only three studies that estimate the distributional effects of implementing a UBI in Brazil: Siqueira (2001), IMF (2017), and Rigolini et al. (2020). The latter two are in fact cross-country comparative studies and do not attempt to provide any country-specific result on the desirability or feasibility of a UBI. In particular, IMF (2017) does not consider any compensating scheme to make the UBI reform revenue neutral, while Rigolini et al. (2020) simulates changes in the social assistance system only, with all existing contributory benefits and pensions excluded from the analysis. By its turn, Siqueira (2001) provides a detailed distributive analysis of hypothetical UBI reforms in Brazil, but the simulations do not allow for any change in the current transfer or tax systems. In these cases, reported fiscal and distributional effects can significantly diverge from those derived from a more comprehensive approach which considers the introduction of a UBI along with reforming the existing tax and transfer systems.

The structure of the paper follows. Section 2 outlines the details of the alternative UBI systems simulated in this study and briefly describes the simulation method. Section 3 examines the fiscal implications of the reforms. The distributional effects of each reform are analysed in Section 4. Section 5 concludes.

---

[6] In OECD countries, according to MF (2017), public transfers account for around 21% in average of total household income.

## 2. UBI schemes analysed and method

As mentioned above, three hypothetical UBI schemes have been simulated. The first scheme considered (Scheme 1) combines a uniform payment of a basic income to every individual in society with a flat rate income tax on all other incomes, from the first real. Such a system, usually referred to in the literature as 'basic income/flat tax proposal' (see, for instance, Atkinson 1995), is equivalent in terms of distributional impact to the Negative Income Tax (NIT) proposed by Milton Friedman (1962).[7]

In our simulations, existing (contributory and non-contributory) pension benefits are reduced by the amount of the basic income and all other cash benefits are totally replaced by the basic income. On the revenue side of the budget, the current personal income tax and employee social security contributions are abolished. The rate of the new income tax is calculated to ensure that the reform is 'budget neutral', in the sense that increases in net spending are matched by increases in (net) tax revenue, so that the budget deficit is not exacerbated.

Some advocates of UBI believe that the benefit level should be set at an amount large enough to ensure a basic level of income security for everyone, including those without any other source of income. The national poverty line and the median income are often taken as references. In our simulations, the UBI is set at the level of the poverty line suggested by the World Bank for upper-middle-income countries, which is US$5.50 a day. This was equivalent to 51% of the Brazilian per capita median disposable income in 2017 (our reference year).[8]

The second scheme simulated differs from Scheme 1 in that the level of the basic income varies according to the age of the recipient: a standard amount equal to the poverty line is paid to working age adults (18 – 64 years), half this amount is the basic income paid to children (under 18 years), and double the standard amount is paid to elderly people (65 and over).[9] The basic idea here is to enhance fiscal and political feasibility with respect to Scheme 1, as under Scheme 2 the net cost of UBI is expected to be lower, particularly to pensioners. By its turn, the third scheme considered differs from Scheme 2 in that, the income tax has a lower marginal rate on incomes below a certain threshold. This lower rate is set at 20% and it is applied on income levels that are lower than twice the median per capita household gross income.[10]

---

[7] The two schemes differ in the way they are implemented. Under the NIT most individuals receive part or the whole of the basic income grant in the form of tax exemptions.

[8] In 2017, this poverty line corresponded to R$406 per month, equivalent to 43% of the legal minimum wage, as well as of the basic pension paid by the Brazilian social security system in the same year.

[9] In 2017, 65 was the standard statutory retirement age for males in Brazil (although some regimes permitted retirement much earlier).

[10] In 2017, the monthly median per capita household gross income was R$850.

All simulations are performed using a static tax-benefit microsimulation model, BRAHMS (*Brazilian Household Microsimulation System*), specially built to incorporate key features of the Brazilian tax-benefit system.[11] A microsimulation model is a computational programme that calculates tax paid and transfers received by individuals/households in a nationally representative sample of the population. The model takes into account the interaction among the different policy instruments built into the tax-benefit system, and it is thus particularly suitable to evaluating the distributional and budgetary impact of tax and benefit reforms. The particular version of the microsimulation model used in this study is based on the household survey *Pesquisa Nacional por Amostra de Domicílios Contínua* (PNADC) for the year 2017, carried out by the Brazilian Institute of Geography and Statistics.[12] As the model is static, the simulations only estimate first-round effects and do not consider behavioural responses.[13]

The basic microsimulation outcome we are concerned with is the *disposable income* of each household under the existing tax-transfer system and under each UBI reform. Changes in disposable income at the household level determine the distributional effects of the reform and, on the aggregate, they explain the impact on fiscal variables.

**3. Fiscal effects**

In this first section of the results some aggregates are computed that could help to determine the financial feasibility of the UBI schemes defined above. These are presented in Table 1, which shows in the first line household initial income, that is, income before tax and government transfers. Then transfer and tax aggregates are presented, followed by household disposable income, defined as income after taxes and transfers. Table 1 also shows the income tax rates calculated as required to ensure that the reforms are budget neutral, as well as the reduced rate in Scheme 3.

The total amount of transfers paid out by the Brazilian social security system in 2017 which are taken into account in this study was 804 billion reais. This corresponded to 12.2% of GDP and 26.6% of total household disposable income in that same year. Pension benefits (contributory and non-contributory) accounted for 89.2% of these cash transfers. The other (non-pension) transfers are essentially comprised of the unemployment benefit, the *Bolas Familiar* (Family Grant) conditional cash transfer, and in-work benefits (family wage and wage bonus). Looking at the revenue aggregates, in 2017 the personal income tax and employee social security contributions together amounted to R$357 billion, equivalent to 5.4% of GDP and to 16.8% of total tax revenue that same year.

---

[11] BRAHMS is a proprietary model. For its details, see Immervoll, Levy, Nogueira, O'Donoghue, and Siqueira (2006).

[12] PNADC microdata is publicly available and can be accessed at IBGE's PNADC home page, https://www.ibge.gov.br/estatisticas/sociais/rendimento-despesa-e-consumo/17270-pnad-continua.html?=&t=downloads

[13] More details on the essential features of the microsimulation model used in this study are provided in Immervoll et al. (2009).

**Table 1.** Budgetary Effects (billions of reais/year)

| Incomes, transfers and taxes | Current system (2017) | Scheme 1 | Scheme 2 | Scheme 3 |
|---|---|---|---|---|
| Initial (market) income | 2,571 | 2,571 | 2,571 | 2,571 |
| Current transfers | 804 | | | |
|    Pensions | 717 | | | |
|    Others | 87 | | | |
| Current tax revenue | 357 | | | |
|    Personal income tax | 192 | | | |
|    Employee social security contribution | 165 | | | |
| Current disposable income | 3,018 | | | |
| | | | | |
| UBI gross cost | | 1,009 | 969 | 969 |
| Reduction in current transfers | | 251 | 335 | 335 |
| UBI net cost | | 758 | 634 | 634 |
| | | 553 | 469 | 469 |
| Tax revenue under UBI | | 1,115 | 991 | 991 |
| Disposable income under UBI | | 3,018 | 3,018 | 3,018 |
| | | | | |
| Income tax rate under UBI (%) | | | | |
|    Flat/Standard | | 35.7 | 32.6 | 47.5 |
|    Reduced | | - | - | 20.0 |

*Source:* Authors' calculations using PNADC 2017 and BRAHMS model.

The gross cost of the UBI is around R$1 trillion (about 15% of GDP in 2017) in Scheme 1, and only slightly lower (R$969 billion) under Schemes 2 and 3. However, the elimination of the current non-pension benefits along with the downward adjustment of pensions offset nearly 25% of the gross cost of the UBI under Scheme 1, and nearly 35% under Schemes 2 and 3. Note that the total removal of the existing benefits would enable the government to offset about 80% of the UBI gross cost. It can be verified that, as intentioned by the microsimulation model, total disposable income after each UBI reform matches the current disposable income.

The flat tax rates that ensure the budget neutrality of Schemes 1 and 2 are respectively 35.7% and 32.6%. These rates are lower than the marginal tax rate on some higher income individuals under the 2017 tax system, which reaches 38.5%, taking the personal income tax and employee social security contribution together. However, in Scheme 3, where we establish the rate of 20% on lower incomes, the marginal tax rate on higher incomes must be 47.5% for revenue neutrality.

Although total disposable income before and after each reform is equal, at the household level the UBI reforms produce changes in disposable income that vary substantially across income groups, both in magnitude and direction. The resulting distributional effects are examined in the next section.

## 4. Distributional effects

This section shows the changes that each UBI scheme would bring about in the distribution of income by looking at poverty and inequality summary indicators, as well as by looking at the patterns of household gains and losses across income groups. Per capita household disposable incomes are used to derive these indicators. These distributional effects are crucial to an assessment of the social desirability of the UBI reforms, and besides they can shed light on political feasibility.

### 4.1. Poverty and inequality indicators

Table 2 shows the head count indicator of poverty – estimated for the whole population and by age group – under the current (2017) tax-transfer system and after each alternative UBI reform. It can be seen that under the existing tax-transfer system the proportion of the total population in poverty is 23.5%. However, poverty among children is much higher, at 39.7%, whereas the corresponding indicator for old age people is 3.2%. As mentioned in the introduction, this reflects the dual nature of the existing social protection system with quite generous pension regimes but relatively small transfers to low income families with children.

Under Scheme 1, by design, poverty is eliminated, as the (non-taxable) basic income paid to every individual is set at the level of the full value of the poverty line. Scheme 2, where the level of the basic income varies according to the age group of the recipient (100% of the poverty line to working age adults, half this amount to children, and twice the poverty line to the elderly), reduces overall poverty by nearly two thirds, and child poverty by 55.7%, while old age poverty is virtually eliminated. The impact of Scheme 3 on poverty is similar to Scheme 2, but a little more pronounced given the reduced tax rate on lower incomes. Under the former the poverty rate among children would fall by 62.2%.

**Table 2: Effects on Poverty and Inequality**

| Inequality and poverty indices | Current system (2017) | Scheme 1 | Scheme 2 | Scheme 3 |
|---|---|---|---|---|
| **% of individuals in poverty** | | | | |
| Total population | 23.5 | 0.0 | 8.0 | 6.8 |
| % reduction | - | 100.0 | 66.0 | 71.1 |
| Children (< 18) | 39.7 | 0.0 | 17.6 | 15.0 |
| % reduction | - | 100.0 | 55.7 | 62.2 |
| Working age (18 - 64) | 20.5 | 0.0 | 5.5 | 4.7 |
| % reduction | - | 100.0 | 73.2 | 77.1 |
| Old age (>= 65) | 3.2 | 0.0 | 0.3 | 0.2 |
| % reduction | - | 100.0 | 90.1 | 93.7 |
| **Gini coefficient of inequality** | 0.506 | 0.377 | 0.408 | 0.373 |
| % reduction | - | 25.5 | 19.4 | 26.3 |

*Source:* Authors' calculations using PNADC 2017 and BRAHMS model.

To summarize the potential impact of the UBI reforms on income inequality, we estimate the Gini coefficient before and after the introduction of each scheme.[14] The last two lines of Table 2 show that any UBI scheme simulated would deliver a substantial reduction in inequality. Under Schemes 1 and Scheme 3 the Gini coefficient would fall by 25.5% and 26.3%, respectively. Scheme 2 is the less progressive, but the estimated reduction in inequality, 19.4%, would still be pronounced. It may be interesting to note that by reducing the Gini from nearly 0.51 to around 0.37, reforms 1 and 2 would bring Brazil´s inequality, in terms of the Gini coefficient, much closer to the OECD average of 0.31, and on pair with the United Kingdom´s 0.36.[15]

### 4.2. Distributional effects in terms of winners and losers

Another way of inspecting the distributional outcomes is to define them in terms of gains and losses at the household level. Net gains occur when the UBI payment outweigh reductions in existing benefits and increased tax for a given household, and vice versa for net losses. For the purposes of this exercise individuals are grouped in income deciles based on the distribution of per capita household disposable income under the current tax-transfer system.

---

[14] The Gini coefficient is a standard measure of inequality, which varies from 0 to 1. The closer to 1 the coefficient the higher the level of inequality.

[15] See OECD Data, https://data.oecd.org/inequality/income-inequality.htm.

Each table in this section shows, for a given simulated reform, the percentage of winners and losers with respect to the current (2017) scenario, current average per capita household disposable income, and average per capita gains and losses, by decile group. It should be noted that average gains and losses are computed among losing and gaining households. This section also provides a graphical representation of the redistribution pattern associated with each reform, where average gains and losses are presented as proportion of current household disposable income.

Table 3 shows the pattern of gains and losses associated with the implementation of Scheme 1. It can be seen that virtually everyone that (under the existing social protection system) is among the poorest 40% of the population and 90% of those in the fifth income decile would have their incomes uplifted by the introduction of Scheme 1. The proportion of losers overcome the proportion of winners only from the seventh highest decile and higher. On the whole, 64% of the population improve their situation after the introduction of Scheme1.

**Table 3: Percentage of winners and losers with respect to the 2017 situation, with average per capita household disposable income and average gain and loss – Scheme 1**

| Deciles | Winners % | Baseline income R$/month | Gain R$/month | Losers % | Baseline income R$/month | Loss R$/month |
|---|---|---|---|---|---|---|
| 1 | 100 | 119 | 341 | 0 | - | - |
| 2 | 100 | 291 | 273 | 0 | - | - |
| 3 | 100 | 430 | 222 | 0 | - | - |
| 4 | 98 | 564 | 168 | 2 | 558 | 79 |
| 5 | 90 | 716 | 134 | 10 | 715 | 41 |
| 6 | 76 | 892 | 99 | 24 | 921 | 93 |
| 7 | 45 | 1,077 | 65 | 55 | 1,059 | 149 |
| 8 | 27 | 1,307 | 39 | 73 | 1,381 | 129 |
| 9 | 3 | 1,793 | 45 | 97 | 1,933 | 255 |
| 10 | 1 | 3,260 | 48 | 99 | 4,776 | 788 |
| **All deciles** | **64** | **565** | **194** | **36** | **2,360** | **342** |

*Source:* Authors' calculations using PNADC 2017 and BRAHMS model.

It is easier to assess the magnitude of the gains and losses by looking at Figure 1. The average gain in the lowest decile is close to three times as high as the current average disposable income. For the next three deciles the average gain is also substantial, equivalent to 94%, 52% and 30% of current average disposable income, respectively. On the other hand, while almost all individuals in the top two quintiles are net losers, the average losses are not so expressive, being around 13% in the ninth decile and around 16% in the highest decile.

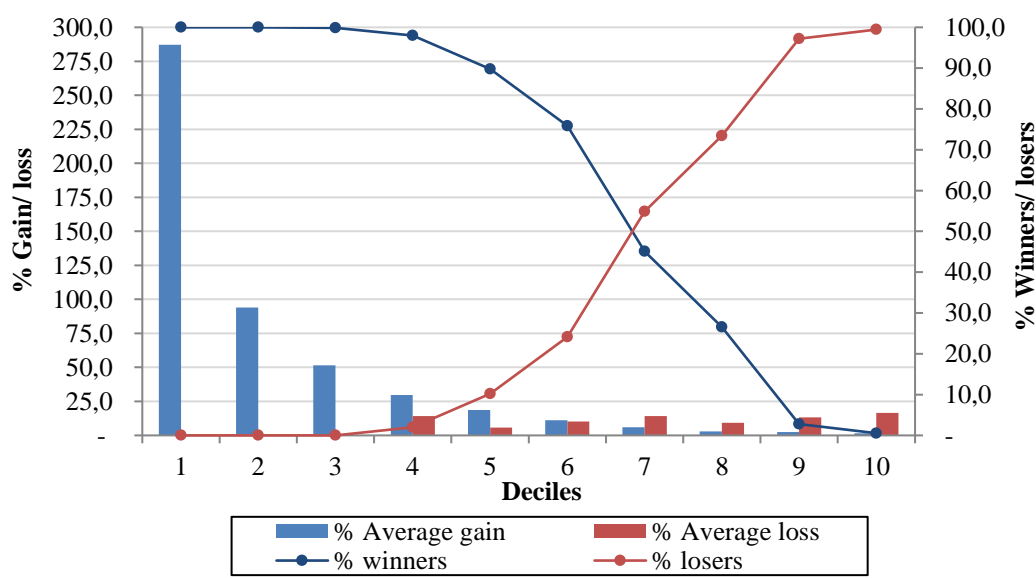

**Figure 1: Percentage of winners and losers and percentage change in household disposable income, by income decile – Scheme 1**

*Source:* Authors' calculations using PNADC 2017 and BRAHMS model.

Table 4 and Figure 2 show the results associated with Scheme 2. Although the distribution of gainers and losers among income groups is very similar to Scheme 1, gains and losses are smaller in average. Yet the gains for individuals at the bottom of the income distribution are still substantial. In the case of the poorest 10%, average disposable income more than doubles, and in the second decile the average gain is equivalent to more than two thirds of the current disposable income. In the ninth and the highest deciles the average losses are even lower than in Scheme 1, respectively 11% and 13%.

**Table 4: Percentage of winners and losers with respect to the 2017 situation, with average per capita household disposable income and average gain and loss – Scheme 2**

| Deciles | Winners % | Baseline income R$/month | Gain R$/month | Losers % | Baseline income R$/month | Loss R$/month |
|---|---|---|---|---|---|---|
| 1 | 100 | 119 | 253 | 0 | - | - |
| 2 | 100 | 290 | 199 | 0 | - | - |
| 3 | 99 | 430 | 168 | 1 | 446 | 38 |
| 4 | 96 | 564 | 134 | 4 | 562 | 60 |
| 5 | 91 | 715 | 107 | 9 | 723 | 52 |
| 6 | 77 | 894 | 82 | 23 | 918 | 64 |
| 7 | 46 | 1,074 | 60 | 54 | 1,062 | 88 |
| 8 | 29 | 1,317 | 43 | 71 | 1,380 | 101 |
| 9 | 5 | 1,824 | 51 | 95 | 1,935 | 206 |
| 10 | 2 | 4,353 | 72 | 98 | 4,776 | 627 |
| **All deciles** | **65** | **584** | **147** | **35** | **2,363** | **269** |

*Source:* Authors' calculations using PNADC 2017 and BRAHMS model.

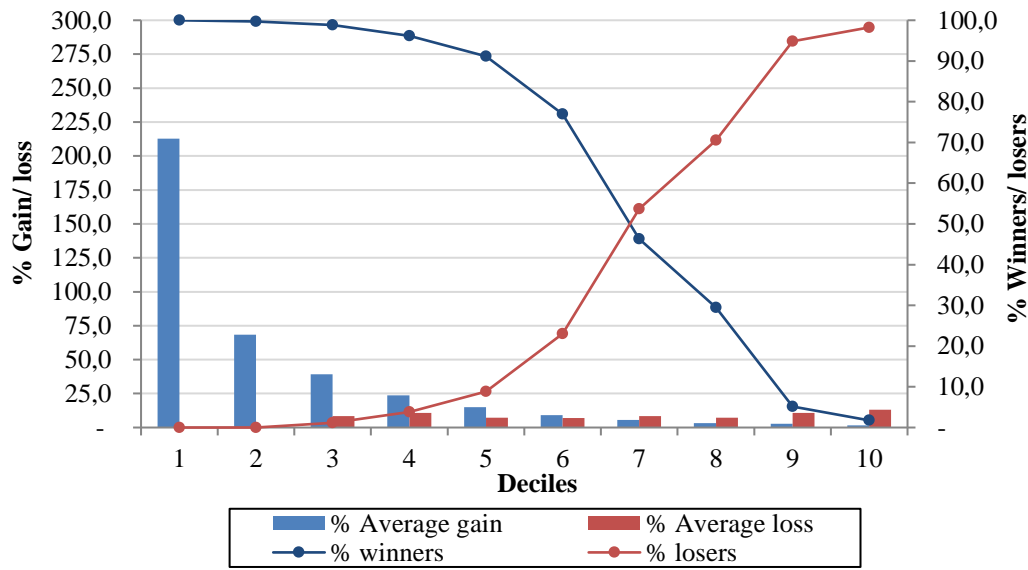

**Figure 2: Percentage of winners and losers and percentage change in household disposable income, by income decile – Scheme 2**

*Source:* Authors' calculations using PNADC 2017 and a tax-benefit microsimulation model.

Table 5 and Figure 3 show that almost all individuals in the half bottom of the income distribution would have their net incomes considerably increased with the implementation of Scheme 3. The percentage of losers is above the percentage of winners only in the top two quintiles. On the whole, 72% of the population would benefit from Scheme 3, compared to 64% and 65% in Schemes 1 and 2, respectively. At the same time, the magnitude of gains and losses would be roughly the same as in Schemes 1 and 2. On the other hand, the average loss to individuals in the highest decile would increase to the equivalent of nearly 23% of current disposable income.

**Table 5: Percentage of winners and losers with respect to the 2017 situation, with average per capita household disposable income and average gain and loss – Scheme 3**

| Deciles | Winners % | Baseline income R$/month | Gain R$/month | Losers % | Baseline income R$/month | Loss R$/month |
|---|---|---|---|---|---|---|
| 1 | 100 | 119 | 263 | 0 | - | - |
| 2 | 100 | 290 | 229 | 0 | - | - |
| 3 | 99 | 430 | 212 | 1 | 442 | 48 |
| 4 | 98 | 565 | 186 | 2 | 548 | 76 |
| 5 | 96 | 715 | 168 | 4 | 727 | 55 |
| 6 | 88 | 895 | 147 | 12 | 929 | 58 |
| 7 | 63 | 1,077 | 124 | 37 | 1,051 | 66 |
| 8 | 56 | 1,338 | 105 | 44 | 1.392 | 92 |
| 9 | 16 | 1,769 | 72 | 84 | 1,960 | 209 |
| 10 | 0 | 2,493 | 123 | 100 | 4,769 | 1,075 |
| **All deciles** | **72** | **639** | **185** | **28** | **2,669** | **468** |

*Source:* Authors' calculations using PNADC 2017 and BRAHMS model.

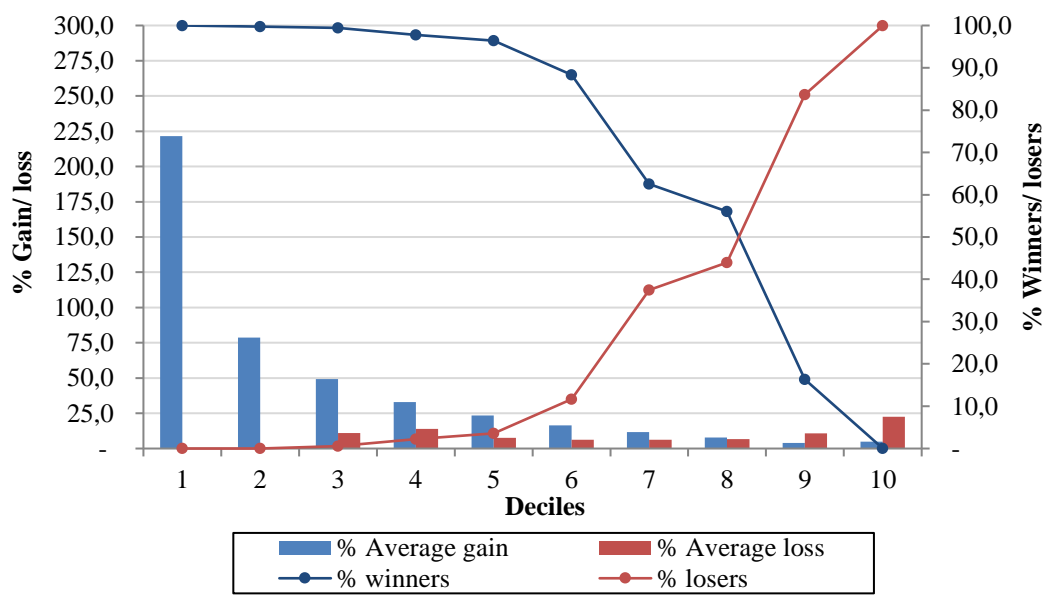

**Figure 3: Percentage of winners and losers and percentage change in household disposable income, by income decile – Scheme 3**

*Source:* Authors' calculations using PNADC 2017 and a tax-benefit microsimulation model.

## 5. Conclusion

In this paper, we have used a static microsimulation model to analyse the fiscal and distributional effects associated with the implementation in Brazil of alternative reforms which combine a UBI with a simplified tax structure. We examined the impact of each hypothetical reform on poverty and inequality, and looked at the distribution of winners and losers across income groups.

The results show that a UBI can be an extremely effective strategy to reduce poverty and inequality in Brazil. In fact, the great majority of people would have their incomes uplifted by the introduction of any of the simulated reforms, with substantial gains concentrated at the bottom of the income distribution, whereas losses would be relatively small in average and concentrated in the top income groups.

These findings suggest that the introduction of a UBI in Brazil is politically and economically viable. Thus, Brazil appears as an exception to the so-called *demand-capacity paradox* identified by UBI analysts, according to which countries where the introduction of a UBI would benefit the largest number of people are the countries with the most limiting fiscal conditions to implement it (see, for example, Wispelaere and Yemtsov, 2020).

It should be remarked that the purpose of this paper was to investigate the social desirability and economic feasibility of a UBI in Brazil, and not to propose a specific design for practical implementation. More work is needed to explore additional ways of financing, including consumption taxation, the elimination of some inefficient ill-targeted programs, and the abolition of numerous regressive fiscal subsidies. The revival of the political and public debate on UBI proposals prompted by the coronavirus pandemic makes further research essential.